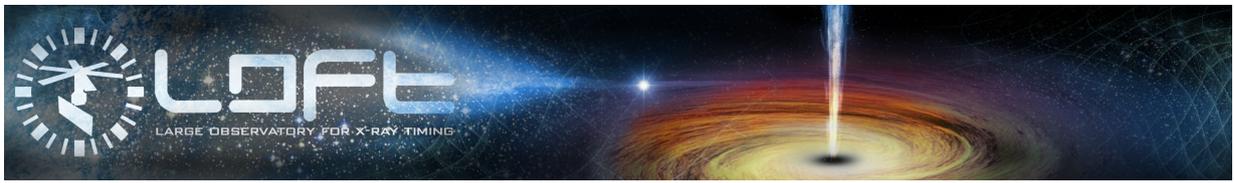

# The physics of accretion-ejection with *LOFT*

White Paper in Support of the Mission Concept of the Large Observatory for X-ray Timing

## Authors

P. Casella[1], R. Fender[2], M. Coriat[3], E. Kalemci[4], S. Motta[2], J. Neilsen[5], G. Ponti[6], M. Begelman[7], T. Belloni[8], E. Koerding[9], T.J. Maccarone[10], P.-O. Petrucci[11], J. Rodriguez[12], J. Tomsick[13], S. Bhattacharyya[14], S. Bianchi[15], M. Del Santo[16], I. Donnarumma[17], P. Gandhi[18], J. Homan[19], P. Jonker[20], M. Kalamkar[1], J. Malzac[21], S. Markoff[22], S. Migliari[23], J. Miller[24], J. Miller-Jones[25], J. Poutanen[26], R. Remillard[19], D.M. Russell[27], P. Uttley[22], A. Zdziarski[28]

[1] INAF, Osservatorio Astronomico di Roma, Via Frascati 33, I-00040 Monteporzio Catone, Italy
[2] Department of Physics, University of Oxford, Keble Road, OX1 3RH Oxford, UK
[3] University of Cape Town, Private Bag X3, Rondebosch 7701, South Africa
[4] Faculty of Engineering and Natural Sciences, Sabancı University, Orhanlı-Tuzla, 34956, Istanbul, Turkey
[5] Einstein Fellow, Boston University Department of Astronomy, Boston, MA 02215, USA
[6] Max-Planck-Institut für extraterrestrische Physik, Giessenbachstrasse, 85748 Garching, Germany
[7] JILA, University of Colorado and National Institute of Standards and Technology 440 UCB, Boulder, CO 80309-0440, USA
[8] INAF, Osservatorio Astronomico di Brera, via E. Bianchi 46, 23807 Merate, Italy
[9] Department of Astrophysics/IMAPP, Radboud University Nijmegen, PO Box 9010, 6500 GL Nijmegen, the Netherlands
[10] Department of Physics, Texas Tech University, Box 41051, Lubbock, TX 79409, USA
[11] UJF-Grenoble 1/CNRS-INSU, IPAG, UMR 5274, 38041 Grenoble, France
[12] Laboratoire AIM, Irfu/Service d'Astrophysique, CEA-Saclay, 91191 Gif-sur-Yvette Cedex, France
[13] Space Sciences Laboratory, 7 Gauss Way, University of California, Berkeley, CA 94720-7450, USA
[14] Department of Astronomy and Astrophysics, Tata Institute of Fundamental Research, 1 Homi Bhabha Road, Mumbai 400005, India
[15] Dipartimento di Matematica e Fisica, Università degli Studi Roma Tre, Via della Vasca Navale 84, 00146 Roma, Italy
[16] Istituto Nazionale di Astrofisica, IASF Palermo, Via U. La Malfa 153, 90146 Palermo, Italy
[17] INAF-Istituto di Astrofisica e Planetologia Spaziale, Via Fosso del Cavaliere 100, 00133 Rome, Italy
[18] School of Physics & Astronomy, University of Southampton, Highfield, Southampton, SO17 1BJ, UK
[19] Kavli Institute for Astrophysics and Space Research, MIT, 70 Vassar Street, Cambridge, MA 02139, USA
[20] SRON, Netherlands Institute for Space Research, Sorbonnelaan 2, 3584 CA Utrecht, the Netherlands
[21] Universite de Toulouse; UPS-OMP; IRAP; Toulouse, France
[22] Astronomical Institute Anton Pannekoek, University of Amsterdam, Science Park 904, 1098XH Amsterdam, the Netherlands
[23] Department of Astronomy and Meteorology & Institute of Cosmic Science, University of Barcelona, Martí i Franquès 1, 08028 Barcelona, Spain
[24] Department of Astronomy, University of Michigan, Ann Arbor, MI 48109, USA
[25] International Centre for Radio Astronomy Research, Curtin University, GPO Box U1987, Perth, WA 6845, Australia
[26] Astronomy Division, Department of Physics, PO Box 3000, 90014 University of Oulu, Finland
[27] New York University Abu Dhabi, PO Box 129188, Abu Dhabi, UAE
[28] Centrum Astronomiczne im. M. Kopernika, Bartycka 18, 00-716 Warszawa, Poland

**Preamble**

The Large Observatory for X-ray Timing, *LOFT*, is designed to perform fast X-ray timing and spectroscopy with uniquely large throughput (Feroci et al., 2014). *LOFT* focuses on two fundamental questions of ESA's Cosmic Vision Theme "Matter under extreme conditions": what is the equation of state of ultra-dense matter in neutron stars? Does matter orbiting close to the event horizon follow the predictions of general relativity? These goals are elaborated in the mission Yellow Book (http://sci.esa.int/loft/53447-loft-yellow-book/) describing the *LOFT* mission as proposed in M3, which closely resembles the *LOFT* mission now being proposed for M4.

The extensive assessment study of *LOFT* as ESA's M3 mission candidate demonstrates the high level of maturity and the technical feasibility of the mission, as well as the scientific importance of its unique core science goals. For this reason, the *LOFT* development has been continued, aiming at the new M4 launch opportunity, for which the M3 science goals have been confirmed. The unprecedentedly large effective area, large grasp, and spectroscopic capabilities of *LOFT*'s instruments make the mission capable of state-of-the-art science not only for its core science case, but also for many other open questions in astrophysics.

*LOFT*'s primary instrument is the Large Area Detector (LAD), a $8.5\,\mathrm{m}^2$ instrument operating in the 2–30 keV energy range, which will revolutionise studies of Galactic and extragalactic X-ray sources down to their fundamental time scales. The mission also features a Wide Field Monitor (WFM), which in the 2–50 keV range simultaneously observes more than a third of the sky at any time, detecting objects down to mCrab fluxes and providing data with excellent timing and spectral resolution. Additionally, the mission is equipped with an on-board alert system for the detection and rapid broadcasting to the ground of celestial bright and fast outbursts of X-rays (particularly, Gamma-ray Bursts).

This paper is one of twelve White Papers that illustrate the unique potential of *LOFT* as an X-ray observatory in a variety of astrophysical fields in addition to the core science.

# 1 Summary

Understanding the dynamics and stability of gas flows around compact objects such as black holes (BH) remains one of the most pressing problems in high-energy astrophysics. A full understanding of such flows is necessary to build a complete picture of X-ray binaries and active galactic nuclei (AGN). This includes their radiative output, the formation of (sometimes superluminal) jets and the launch of mass-loaded accretion disk winds. The radiation as well as the kinetic output of accreting supermassive BHs probably plays an important role in regulating the growth of the biggest galaxies in our Universe. To fully understand AGN feedback, one has to understand the interplay between the three main outputs of accreting objects: the radiation, winds, and jets. Notwithstanding the large differences between stellar-mass and super-massive BHs, the physics at play near the BH is expected to be substantially the same, albeit scaled with the different length scales, time scales and energy densities at work. Studies of BH accretion and outflows have exploited the differences between the two BH classes through complementary approaches. The number of photons per characteristic time scale is typically several orders of magnitude higher in AGN, which has allowed us to observe the detailed physics at play in individual "events". On the other hand, stellar-mass BHs have much shorter characteristic time scales, allowing us to observe a large number of cycles, providing a good picture of the average behaviour of each physical process and control for the effects of unknown parameters such as distance, mass or spin. The ongoing improvement in detector efficiency, together with the increase in telescope area, are now bridging the gap across the BH-mass range, as the fastest characteristic time scales start to become accessible even for stellar-mass BHs. Many basic questions remain unanswered, with unknowns including the geometry and the radiative efficiency of the accretion flow, the physical mechanism triggering the transitions between different accretion regimes, the particle composition of jets, the amount of mass carried by the winds, and the launching and powering mechanisms of jets and winds. *LOFT* will revolutionise this field. Thanks to the huge effective area of the LAD, its CCD-class spectral resolution, and to the outstanding monitoring performances of the WFM, *LOFT* will:

- measure the X-ray spectral continuum with unprecedented accuracy, allowing us to study the detailed evolution of the accretion disk and the Comptonizing medium on the time scale of observed transitions: in a few seconds during hard-to-soft bright transitions, a few hours during the soft-to-hard faint transitions;

- detect X-ray absorption and emission lines from winds and jets with outstanding statistics, allowing us to identify the moments when the wind and the jet turn on/off, and to track the evolution of their physical properties (including density, ionisation, velocity, baryonic content) on time scales as short as a few seconds and a few hours during the bright and the faint transitions, respectively;

- identify and pinpoint each fast transition with the WFM, to establish a crucial component of the upcoming multi- wavelength legacy;

- carry out all these breakthroughs at the same time as unprecedented measurements of the structure of the inner accreting and emitting regions, to link changes in jet, wind or radiative output directly to changes in the central engine which drives them.

It is important to note that *LOFT* will perform most of these measurements with the observations already planned to match the Core Science requirements.

# 2 Introduction

Accretion onto BHs (and to a comparable degree, neutron stars) is the most efficient source of energy in the contemporary Universe, powering both X-ray binaries (XRBs) and AGN. The energy liberated through accretion emerges in the form of radiation or kinetic energy, that can either be collimated (so-called jets) or uncollimated (winds). The interplay between these three components is not yet understood, neither in stellar-mass nor in supermassive BHs. While jet ejection events, for example, can be studied in detail in AGN, as their relative time-scale (set by their gravitational radius) is longer, the evolution of an accreting BH under changes of the accretion rate can be better-studied in XRBs. The luminosity of accreting stellar-mass BHs can vary by as much as 8 orders of magnitude between quiescence and peak outburst within a couple of months. As it changes its luminosity, an XRB evolves through a sequence of accretion states. In particular, two canonical states are seen – a "hard" state characterised by Comptonized hard X-ray emission, high amplitude X-ray variability and a compact radio jet, and a "soft" state characterised by strong X-ray thermal emission, low amplitude X-ray variability and an equatorial wind. Transitions between these canonical states (Begelman & Armitage, 2014) proceed through a variety of intermediate and extreme states, in which the accretion energy budget redistributes itself among the various components on time scales as short as a few seconds, as evident from X-ray variability (see, e.g., Belloni, 2010). These sources are thus ideally suited for studying the formation of jets and winds, which appear and subsequently disappear during each outburst cycle with complex phenomenology on a broad range of time scales. We will thus discuss the three different outputs of an accreting BH: radiation, wind and the jets in view of the possibilities *LOFT* will offer in the future. *LOFT* will for the first time be able to detect the disc wind routinely, and thus offer in a single observation information about both the radiative part as well as the mechanical energy in the wind. The proposed launch date and lifetime of *LOFT* correspond nicely to the period of initial surveys with the phase 1 of the Square Kilometre Array. Due to its increased sensitivity it will then be possible to obtain dense coverage of the radio light-curve, allowing also for a well studied jet component. As we will argue below, *LOFT* will open a new era of research into the coupling between accretion and ejection.

## 3 Getting close to black holes: unlocking the origin of the most powerful jets and winds

In what follows, we detail the breakthroughs that *LOFT* will make in tackling the physics of black hole jets, winds and the radiative emission. But first, it is important to stress that all these results will be obtained *at the same time* as the unprecedented measurements of dynamics and structure of the inner accretion flow which form the *LOFT* core science, aimed at testing and quantifying the effects of strong-field gravity (SFG; see http://sci.esa.int/loft/53447-loft-yellow-book). These measurements will use high-throughput spectral-timing of rapid variability to carry out X-ray reverberation mapping on sub-gravitational-radius scales, and Doppler tomography of QPOs to measure the dynamics of the accretion flow. This suite of independent techniques will provide accurate measurements of black hole spin, the disc inner radius and the geometry and location of the X-ray power-law emitting corona. Moreover, these measurements will be obtained throughout the range of states seen in X-ray binary outbursts, notably the most luminous transitional states when the most powerful jets and winds switch on and off and the X-ray emission shows the strongest spectral evolution.

Thus, *LOFT*'s SFG measurements will allow the detailed changes in structure and geometry of the inner flow and emitting regions during state transitions to be mapped directly on to the properties of the jet or wind and the emission spectrum, *in real time as the system evolves*. Energetically, we know that the most powerful and energetic outputs must have their origin in these inner regions. The same phenomena can be seen in AGN and must evolve on much longer time-scales which we cannot observe directly. *LOFT* studies of BH X-ray binaries will therefore allow an unparalleled insight into the physical causes of accretion-powered jets and winds at accretion rates up to the Eddington limit, providing the crucial (and currently missing) information for modelling accretion and feedback in their supermassive cousins, which may ultimately be responsible for re-shaping entire galaxies and galaxy clusters.

## 4 The radiative output: the X-ray continuum

The X-ray continuum emission from accreting BHs has been studied and modelled for decades. In the hard state, the spectral energy distribution is well-modelled by a power law with a photon index of about 1.7 and a cutoff at a few hundred keV, while in the soft state, thermal emission from the geometrically thin accretion disk dominates, and the power law tail is steeper, with no trace of a cutoff out to at least 1 MeV (Grove et al., 1998). These spectra are broadly interpreted as the result of Comptonization from a combination of (inflowing or outflowing) thermal and non-thermal electrons (see, e.g., Markoff et al., 2001, 2005; Ibragimov et al., 2005; Poutanen & Vurm, 2009). Despite the wealth of observational and theoretical efforts in understanding this emission, many open questions remain on the geometry and energy density of the Comptonizing electron populations, mostly because of the degeneracies between different theoretical models. Time-resolved spectral studies suggest that the Comptonizing medium is inhomogeneous, with multiple electron populations (for example radially stratified in temperature/optical depth, or outflowing) contributing to the observed complex Comptonized emission. The statistics of the LAD data is such that it will be possible to deconvolve the continuum from its reflection as a function of radius, testing theoretical models and measuring the physical parameters of the multiple electron populations.

### 4.1 Bright and fast: physical measurements in a few seconds

As the geometry and thermal properties of the accretion flow are expected to be drastically different in the hard and in soft states, a promising way to discriminate between models is to observe in detail transitions between these two accretion states. The brightest of such transitions, out of the hard state, are associated with the ejection of powerful transient relativistic jets (Fender et al., 2009), while the faint transitions back to the hard state during the outburst decay are associated with the reappearance of the compact steady jet. As these jets are powerful, it is reasonable to expect their ejection to have a substantial impact on the accretion flow. Nevertheless, during the bright, fast transitions, the spectral parameters below 10 keV vary little and smoothly, seemingly unaware of the presence/absence/properties of the highly variable jet. Strong and non-monotonic changes are indeed observed in the properties of the hard tail in a few hours or days (Motta et al., 2009), but nothing is known about the spectral variability on time scales as short as those of the observed X-ray transitions, during which quasi-periodic oscillations and broad-band noise appear/disappear in a few seconds. The epochs of these rapid transitions indeed seem to be associated with rapid geometrical reconfigurations of the accretion flow (Motta, 2014), but a proper physical understanding of them has been hampered so far by the limited statistics, which has prevented spectral analysis on such short time scales. The large effective area over a broad energy range of the LAD/*LOFT* will change this, making the spectral evolution accessible on the fast time scales of the transitions. During these hard-to-soft transitions, the timing properties are often observed to change rapidly, with the broad-band noise and quasi-periodic oscillations appearing/disappearing on time scales shorter than 10 seconds (Casella et al., 2004), sometimes multiple times (Motta et al., 2011). Figure 1 shows two example of LAD spectra with very short exposures. The two spectra simulate the expected spectral variability during such extreme transitions, as inferred by long averaged RXTE observations. Exposures as short as 2 seconds will be sufficient to detect significant differences in the spectra, measuring all spectral parameters with a 5% accuracy or better. At the same time, it will be possible to examine on very short time scales the detailed properties of the sub-second oscillations detected before and after transitions. Linking them to the spectral results will allow us to associate them to changes in spectral parameters and to establish their connection with the accretion and ejection flows.

### 4.2 Dim and slow: monitoring during outburst decay

The dim, slow transition back to the hard state during the outburst decay is another crucial epoch to test models and understand accretion physics (Maccarone, 2003). The relevant time scales here are different, as the transition lasts days or weeks, and the fluxes are orders of magnitude fainter. The compact steady jet is known to reappear during this slow transition, although the actual association between the jet properties and the X-ray spectral evolution is still debated (Kalemci et al., 2005, 2013; Dinçer et al., 2014). Despite coordinated efforts to understand the evolution of the inner disk radius as a function of luminosity during the subsequent decay phase (Tomsick et al., 2009; Petrucci et al., 2014; Plant et al., 2014), we have yet to determine how the disk moves away from the BH during decays. Also to be understood is the marked X-ray softening observed in some sources during the decay, which could be due to an evolution of the properties of the Comptonizing medium (geometry, electron temperature and density, seed photons; Sobolewska et al., 2011) or to the presence of a second source of emission (perhaps a jet; Russell et al., 2010) dominating below a critical luminosity. The limited effective area of past and current X-ray satellites, and the consequent need for long exposures, have hampered so far a proper monitoring of this transition. The LAD large effective area will allow us to track the evolution of the spectral properties throughout the transition and early phases of the decay with exposures as short as a few hours. For a 5 mCrab source, a 2 ks LAD observation will provide better constraints than a 40 ks *Suzaku* exposure (as reported by Petrucci et al., 2014) to the parameters of the hard X-ray component and – if the goal of 1.5 keV soft energy limit is reached – even to those of the soft disk component. At similar flux levels, a 5 ks LAD exposure will be enough to constrain the inner disk radius through iron line modelling, while the exposure time should be increased to around 25 ks to obtain the full evolution down to a 0.5 mCrab flux level.

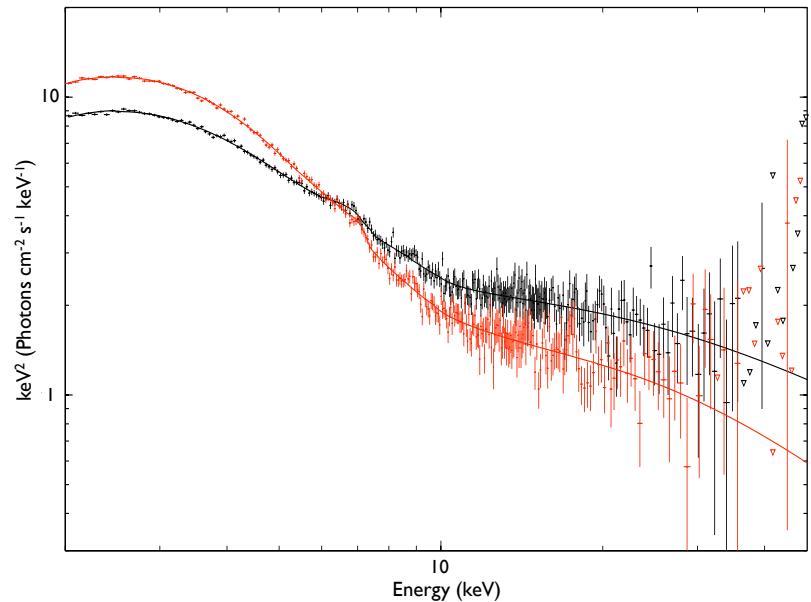

Figure 1: Simulated 2 second exposures LAD/*LOFT* spectra for GX 339−4. The black and red spectra come from observations with and without broadband noise variability, respectively. Data were fitted with an absorbed disk black-body plus power law plus reflection. The reflection component has been corrected for relativistic effects.

### 4.3 The multi-wavelength legacy: solving degeneracies through multi-wavelength timing

In recent years, there has been a tremendous amount of information coming from wavelengths other than X-rays, ranging over the whole electromagnetic spectrum, from radio all the way to the TeV range. Crucially and perhaps not surprisingly, the properties of accreting BHs in these other wavebands have been shown to correlate with those in X- rays. Among these many new observational windows, high time resolution observations at optical-infrared wavelengths have yielded the important discovery of variable non-thermal emission correlated with the X-ray flux on the shortest accessible time scales (Malzac et al., 2004; Gandhi et al., 2010, e.g.). Dedicated fast optical/IR photometers with good quantum efficiency allow us to tackle old questions from a new perspective: the fastest time scales of accretion onto stellar-mass BHs are becoming accessible for study in the synchrotron emission from the same electron population responsible for the X-ray Comptonized emission (e.g., Veledina et al., 2013). At longer wavelengths we can now study synchrotron emission from the electron population outflowing in the relativistic jet (Casella et al., 2010). Such observations are crucial in order to solve X-ray continuum degeneracies, as different physical scenarios – which would otherwise predict very similar X-ray spectral properties – predict extremely different Optical/Infrared/X-ray spectral timing properties. If the first key results have been limited so far by the non optimal OIR technologies and by the available X-ray statistics, the next decade will see significant technological advances. Observations of rapid variability in all wavebands will provide revolutionary new data for the study of accretion and ejection.

In particular, the field will benefit from the combination of the next generation of OIR detectors – such as the Microwave Kinetic Inductance Detectors (MKIDs), which are able to measure the energies of individual optical/IR photons without using filters or gratings, obtaining similar spectral resolutions to those possible with X-ray CCD instruments (e.g., the ARCONS camera; Mazin et al., 2013) – and the large statistical throughput offered by *LOFT*. Equally important will be the expected increased availability of 4–8 m class telescopes, on which to mount such instruments, which will provide the photon count rates needed to study rapid variability. The sum total of all this will result in a twofold field expansion: (a) physical measurements via correlated OIR/X-ray timing observations will be accessible on the time scales of the fastest transitions, helping to reveal the expected geometrical reconfigurations; and (b) it will be possible to perform such measurements, now limited to the brightest fluxes and/or the brightest sources, at order-of-magnitude lower fluxes, allowing full monitoring through the whole outburst evolution of several sources.

## 5 The mechanical output: winds

Although we have learned a great deal about accretion disk winds while monitoring Galactic BHs in the last two decades (see, e.g, Miller et al., 2006; Done et al., 2007; Miller et al., 2012), there are many remaining questions (see also Tombesi et al., 2013, and references therein, for winds in AGN). When do winds first appear in outburst and why; when do they disappear? Do they interact with jets or disrupt jet formation (Neilsen & Lee, 2009)? We know that winds are reliably detected after the transition out of the hard state and the disappearance of the jet (Ponti et al., 2012; Neilsen & Homan, 2012), but the instant of their formation has not been definitively identified. This moment, however, has sweeping implications: if winds originate after the state transition and jet quenching, they cannot possibly play a major role in those processes. Other equally crucial unknowns are the actual launching region, as well as their launching mechanism (are they launched by magnetic processes that tie them to jets?). Also, although their kinetic energy might be overall much smaller than the jet kinetic power or the radiative luminosity, they may carry a large amount of mass away from the accretion flow, possibly influencing the long-term evolution of the outburst or even that of the binary system and of the spin of the accreting object. Finally, the very structure of these outflows has yet to be studied, as well as the influence that variations in the luminosity spectrum may have.

As the relevant physical processes happen on very fast time scales, time-resolved spectroscopy is the key to answering these questions. Thanks to its large effective area, joined with its CCD-class spectral resolution, the LAD will revolutionise studies of winds from accreting BHs. Such winds are known to be ubiquitous in high-inclination BH XRB. The characteristic signature of an accretion disk wind is the Fe XXVI absorption line, with an equivalent width of $\lesssim$35 eV. *LOFT* will detect these strong absorption lines at $3\sigma$ confidence in **as little as 1–2 seconds** (roughly 1000 times faster than Chandra). This combination of sensitivity and time resolution opens up new avenues of inquiry, including absorption line variability as a probe of the turbulent structure of winds. By comparing observed X-ray variability to fluctuations in the ionisation parameter ($\xi = L/nR^2$, where $n$ is the particle number density and $R$ is the distance to the source), *LOFT* will allow us to disentangle the effects of photo-ionisation and the clumpy structure of the wind.

For representative wind parameters ($N_\mathrm{H} = 5 \times 10^{22}$ cm$^{-2}$ and $log\xi = 4.3$), > 6% fractional RMS variations in the luminosity will induce detectable changes in $\xi$; thus, any anomalous ionisation will be used to map the density structure of the wind. Furthermore, simulations show that wind speeds above $\sim 450\,\mathrm{km\,s^{-1}}$ will be reliably measured by *LOFT*. Thus, the LAD will be able to track not only the structure of winds, but also their mass loss rates as functions of the accretion state and the accretion rate for clues to their formation physics and their influence on the disk and the jet.

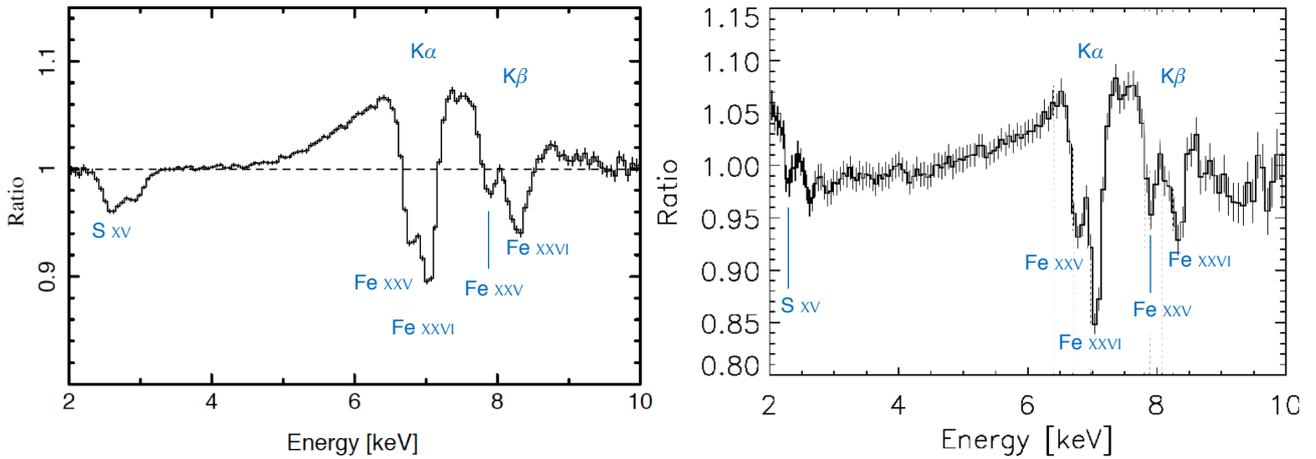

Figure 2: *Left:* Simulated 1 ks LAD exposure of 4U 1630−47 (0.3 Crab, 2–10 keV absorbed flux). The Fe xxv and Fe xxvi lines are clearly resolved into Kα and Kβ. *Right:* the 78 ks *XMM-Newton* observation for comparison (Díaz Trigo et al., 2014).

Complementing the significant advances in our understanding of the structure and variability of winds on short time scales, LOFT will also provide important information about the launching and quenching processes for winds. The LAD will detect weak Fe xxvi lines (equivalent width 7.5 eV) in 1 ks in sources as faint as $2 \times 10^{-10}\,\mathrm{erg\,cm^{-2}\,s^{-1}}$, i.e., more than two orders of magnitude below the present soft state detection. Such short exposures even at low fluxes will allow monitoring observations during both the rising and the decaying phases of BH outbursts. *LOFT* will thus readily reveal the moments when winds first appear and when they eventually quench. Again, these moments are critical for a complete understanding of winds from accreting BHs: is the appearance/disappearance of winds an effect caused by changes in their temperature (thus in their transparency to X-rays), or do they actually start and cease? In order to answer these questions, both a large effective area and a useful spectral resolution are crucial. Indeed, simulations indicate that *LOFT* will be more sensitive than *Athena* to wind absorption lines even at fluxes as low as a few mCrab, with significant detections in sources at least a factor of two fainter. This is because the lines are not intrinsically narrow (widths of ~500 km/s or more are fairly typical), so the improvement due to the larger effective area of *LOFT* dominates over the difference in spectral resolution. At higher fluxes, the effect of pileup might start to play an important role, further decreasing the throughput of *Athena*.

With tight constraints from the LAD on the ionising radiation field throughout these outbursts, *LOFT* will be able track the mass loss rate in winds on the fastest useful time scales from the moment of their formation to their disappearance.

## 6 The mechanical output: jets

In accreting BHs, jets radiate predominantly through processes such as synchrotron and inverse Compton emission, from the radio through X-ray wavebands. Two different types of jet are observed during an outburst: a steady, mildly relativistic jet is detected in the hard state; discrete, very energetic ejections of relativistic matter are observed during the transition toward the soft state, probably associated with rapid changes in the spectral distribution of the hard X-ray emission and with drastic changes in the X-ray timing properties. Among the most hotly currently debated questions there are: how much of the accretion power gets released into the surroundings via these collimated jets? What is their launching mechanism? Are they qualitatively different from winds, or are they an extreme, focused version of the inner part of a wind? What is their composition? Do the jets themselves contribute to the observed X-ray radiation? How and where do jets accelerate particles to very high energies? What is their relation to the variability in the inflow? LOFT will allow us to investigate the role of relativistic jets in combination with the evolution of the characteristics of inflow and wind, helping us to understand further what is physically driving the varied behaviours in these sources.

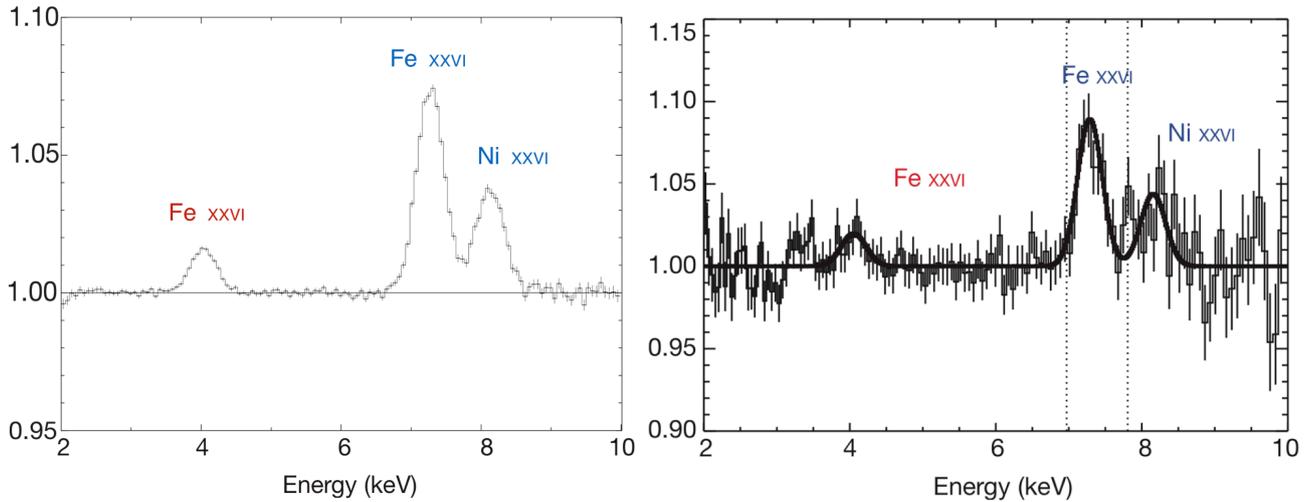

Figure 3: *Left:* simulated 1 ks LAD exposure of 4U 1630−47 (∼2.5 Crab, 2–10 keV absorbed flux). *Right:* the 55 ks *XMM-Newton* burst-mode observation (Díaz Trigo et al., 2013, resulting in 1.6 ks actual exposure), for comparison. The weakest line (the red-shifted Fe xxvi line) which was detected by *XMM-Newton* at a 3.5$\sigma$ level, is detected by *LOFT* at 46$\sigma$ level.

### 6.1 X-ray spectroscopy: the jet baryonic content

The omnipresence of jets in astrophysical systems makes a proper understanding of their physics particularly relevant. Are all these jets produced by the same processes? Since one of the canonical jet formation mechanisms (the Blandford- Znajek, or BZ, mechanism; Blandford & Znajek, 1977), involves the rotation power of a BH, it certainly cannot operate in all systems. An alternative is the Blandford-Payne mechanism (BP; Blandford & Payne, 1982), in which the requisite rotation power comes from a rotating accretion disk. As a consequence, one expects BP jets to be baryonic, since they are magneto-centrifugal outflows from a gaseous reservoir. BZ jets, on the other hand, might be – at least in the early stages after launch – electromagnetically dominated.

Thus it is particularly interesting that accreting stellar-mass BHs appear to produce two different kinds of jets (steady and transient; for reviews see Fender et al., 2004, 2009). Observationally, it remains unclear whether these two types of jets correspond to the two mechanisms described above; a suggested association between the transient jets and the BZ mechanism has been hotly debated (Fender et al., 2010; Narayan & McClintock, 2012; Russell et al., 2013). Different processes may launch jets at different scales or under different conditions, but because a jet's energy requirements are sensitive to its baryon content, these considerations have broad and concrete implications for radio-mode feedback from accreting systems (Fender et al., 1999). The *XMM-Newton* detection of relativistically blue- and red- shifted X-ray emission lines from the black-hole candidate 4U 1630−47 (Díaz Trigo et al., 2013) opened a new perspective on these questions. With an inferred velocity of 0.66$c$, these Fe and Ni lines coincided with the appearance of optically thin radio emission observed by *ATCA*, suggesting the formation of a fast, hot baryonic outflow. An *XMM-Newton*/ATCA observation 3 weeks earlier detected neither radio emission nor X-ray emission lines. If this detection is confirmed, 4U 1630−47 thus becomes the second known source of apparently heavy jets, together with the "odd ball" SS 433 (Margon et al., 1979). In order to understand the full implications of these results for the jet formation mechanism and energetic budget, a large search for baryonic signatures in black hole jets must be carried out. The large effective area and good energy resolution of the LAD will allow tracking of X-ray emissions lines from jets with unprecedented sensitivity and on short time scales. If we detect these emission lines, we will precisely measure their velocities and line widths. The ratio of the line width to the blue-shift is an important diagnostic of the orientation, expansion, and geometry of the jet. Previous *XMM-Newton* observations could only indicate that the lines did not appear to be intrinsically narrow. In addition, the line strengths will constrain the thermal properties and emission measure of the mass-loaded jet. As these quantities evolve, we will correlate them with the X-ray luminosity, the X-ray spectral and timing properties as well as the radio and infrared emission from the jets. The variations of these quantities will provide our first empirical insights into the formation of heavy jets from BHs. Figure 3 (Left panel) shows a LAD simulated spectrum of jets X-ray emissions lines obtained in 1 ks exposure time. An improvement of a factor 15 in signal to noise ratio is achieved with respect to the corresponding 55 ks long *XMM-Newton* observation of 4U 1630−47 (Fig. 3 right panel; Díaz Trigo et al., 2013). For sources with comparable fluxes (which exceed the current flux limitations for *Athena*), exposures as short as 10 seconds will be sufficient to detect the red-shifted Fe XXVI (the faintest) line with *LOFT* at the $4\sigma$ level. Therefore, tracking the jet velocity on minute time scales will become possible with *LOFT*.

## 7 The multi-wavelength legacy and the role of the WFM

The proposed launch date and lifetime of *LOFT* correspond to the period of initial surveys with the Square Kilometre Array (SKA; see skatelescope.org) phase one (SKA1). In the ongoing SKA Science Review Process, Transients and High Energy Astrophysics have been identified as key science drivers for the project, ensuring support for these directions. Although currently undergoing a design re-baselining process, SKA1 is envisaged to have three components, all of which are potentially important contributors to *LOFT* science. SKA1–Low will operate, like LOFAR and MWA at present, at very low frequencies, and will have the advantage of a very large field of view (more than 100 square degrees) and sensitivity to coherent bursts. On the other hand, SKA1-Mid and SKA-Survey will operate in the 1–5 GHz frequency range, and will be two orders of magnitude more sensitive than current facilities, achieving 0.72 microJy in an hour in the 1.4 GHz band (about 40 microJy in one second). Thus, also thanks to the highly flexible scheduling built in, it will be possible to monitor all active X-ray binaries at high cadence. Over the ~5–year lifetime of the SKA1 surveys, it is anticipated that there would be many thousands of high S/N radio observations of X-ray binaries. In optical bands, clear plans exist to make continuous surveys of the sky, as well. The Large Synoptic Survey Telescope (Ivezic et al., 2008) will observe the entire southern sky every three days, down to 24th magnitude. The Zwicky Transient Factory (see www.ptf.caltech.edu/ztf) will monitor the northern sky with more than 300 epochs per year, down to magnitude 20.5 or better. Evryscope (Law et al., 2014) will observe an entire hemisphere of the sky every minute down to magnitude 16.5, with one–hour depth of 19th magnitude. These projects, too, will give wide–field coverage of transients. Furthermore, the Cherenkov Telescope Array (CTA Actis et al., 2011, see www.cta-observatory.org) is the next-generation high energy (from 30 GeV to more than 100 TeV) gamma-ray facility, planned to start full operation by 2020, involving a worldwide collaboration of more than 1000 scientists and about 30 countries. With sensitivity and energy coverage an order of magnitude better than that of H.E.S.S., plus a factor of a few improvement on spatial resolution and field of view, CTA will be run as an observatory, allowing proposals for targeted observations, and including also a ToO mode, with a reaction time of less than a minute. CTA will be the key TeV partner for *LOFT*, for co-monitoring of high-energy emission sites and helping to trace particle acceleration (of both leptons and potentially baryons).